\documentclass[twocolumn,pra,showpacs,superscriptaddress]{revtex4-1}
\usepackage{amssymb}
\usepackage{amsmath}
\usepackage{graphicx}
\usepackage{epsfig}

\begin{document}

\title{Self-sustained emission in semi-infinite non-Hermitian systems at the
exceptional point}
\author{X. Z. Zhang}
\affiliation{School of Physics, Nankai University, Tianjin 300071, China}
\author{L. Jin}
\affiliation{Department of Physics, The Chinese University of Hong Kong, Hong Kong, China.}
\author{Z. Song}
\email{songtc@nankai.edu.cn}
\affiliation{School of Physics, Nankai University, Tianjin 300071, China}

\begin{abstract}
Complex potential and non-Hermitian hopping amplitude are building blocks of
a non-Hermitian quantum network. Appropriate configuration, such as $%
\mathcal{PT}$-symmetric distribution, can lead to a full real spectrum. To
investigate the underlying mechanism of this phenomenon, we study the phase
diagrams of a semi-infinite non-Hermitian systems. They consist of finite
non-Hermitian clusters and semi-infinite leads. Based on the analysis of the
solutions of the concrete systems, it is shown that they can have full real
spectra without any requirements on the symmetry and the wave function
within the leads becomes unidirectional plane waves at the exceptional
point. This universal dynamical behavior is demonstrated as the persistent
emission and reflectionless absorption of wave packets in the typical
non-Hermitian systems containing the complex on-site potentials and
non-Hermitian hopping amplitudes.
\end{abstract}

\maketitle


\section{Introduction}

\label{sec_intro}Non-Hermitian Hamiltonians are often employed to describe
the open systems due to their features of complex-valued energy and
non-preserved particle probability. Recent observations show that a large
families of non-Hermitian Hamiltonians can have all eigenvalues real, if the
loss and gain are set in a balanced manner, being invariant under the
combination of the parity ($\mathcal{P}$) and the time-reversal ($\mathcal{T}
$) symmetry. A parity-time ($\mathcal{PT}$) symmetric non-Hermitian quantum
theory has been well developed as the complex extension of conventional
quantum mechanics \cite{Scholtz,Bender 98,Bender 99,Dorey 01,Bender
02,A.M43,A.M36,A.M,Jones}. Although the condition of the $\mathcal{PT}$
symmetry for a complete real spectrum is weaker \cite{ZXZ}, it still implies
the underlying mechanism can be based on the balance of the loss and gain.
However, such an intuitive consideration of the balance needs to be
investigated precisely. The concept of the balance should not be simply
understood as the conjugate relation of two non-Hermitian subsystems arising
from the $\mathcal{PT}$ symmetry. It can not provide physical explanation to
the following features about exceptional point: (i) The $\mathcal{PT}$
symmetry of the system can not guarantee the balance of the loss and the
gain, or the reality of the energy levels. (ii) The spontaneous symmetry
broken states always appear in pairs. Furthermore, this consideration is
also related to the precise physical significance of the complex potential
and non-Hermitian coupling, which are basic elements for a discrete
non-Hermitian system. On the other hand, the purpose of this investigation
is not only for the fundamental physics,but also for the application in
practice due to the formal equivalence between the quantum Schr\"{o}dinger
equation and the optical wave equation \cite{Ruschhaupt,R. El-Ganainy,K. G.
Makris,Christodoulides,Z. H. Musslimani,S. Klaiman,S. Longhi,H.
Schomerus,LonghiLaser,YDChong,Keya}. Furthermore, the $\mathcal{PT}$
symmetry breaking has been observed in experiments \cite{Guo,Kottos}.

In this paper, we investigate semi-infinite non-Hermitian system without $%
\mathcal{PT}$ symmetry. Based on this, we try to clarify the concept of
balance in the non-Hermitian discrete system in the framework of the quantum
mechanics rather than a phenomenological description. We show an entirely
real spectrum and study the exceptional point of a semi-infinite
non-Hermitian system from the dynamical point of view. We show that the wave
function within the lead becomes a unidirectional plane wave at the
exceptional point. This universal dynamical behavior is demonstrated as the
self-sustained emission and reflectionless absorption of wave packets by two
typical non-Hermitian clusters containing the complex on-site potential and
non-Hermitian hopping amplitude.

This paper is organized as follows. In Section \ref{semi-infinite system} we
analyze the classification of possible solutions and solve two examples to
illustrate our main idea. Section \ref{relation with PT}\ presents the
connection between the semi-infinite systems and $\mathcal{PT}$-symmetric
systems. Section \ref{wavepacket} is devoted to the numerical simulation of
the wave packet dynamics to demonstrate the phenomena of the persistent
emission and reflectionless absorption. Section \ref{sec_summary} is the
summary and discussion.

\section{Semi-infinite system}

\label{semi-infinite system}

The discrete non-Hermitian model, with the non-Hermiticity arising from the
on-site complex potentials as well as the non-Hermitian hopping amplitude,
is a nice testing ground to study the basic features of the non-Hermitian
system not only because of its analytical and numerical tractability but
also the experimental accessibility. In recent years, fundamental aspects of
non-Hermitian continuum systems are studies by using discretization \cite%
{Znojil}, as well as the studies on quantum square wells \cite{a1}. On the
other hand, non-Hermitian quantum models are also investigated, such as
tight-binding systems \cite{Bendix,Liang
Jin,Longhi,Joglekar,Joglekar1,Joglekar2}, spin systems \cite{ZXZ,Korff,T.
Deguchi,Giorgi,ZXZ1}, and strongly correlated systems \cite{H. Zhong,L. Jin}%
. Besides the fundamental features of discrete $\mathcal{PT}$-symmetric
quantum systems, theoretical research on the quantum dynamics and scattering
behaviors in discrete non-Hermitian networks are investigated in a series of
papers \cite{Znojil2008,Stefano,ZXZ1,a2,a3}. In experiment, light transport
in large-scale temporal lattices is studied in $\mathcal{PT} $-symmetric
fiber networks, it is also demonstrated that the $\mathcal{PT}$-symmetric
network can act as a unidirectional invisible media \cite{a4}. Although many
surprising features and possible applications of $\mathcal{PT}$-symmetric
are revealed, they are mostly based on the finite systems. In this paper, we
intend to study the infinite system.

\subsection{Classification of solutions}

Here we consider a semi-infinite lead coupled to a non-Hermitian finite
cluster. The Hamiltonian is written as
\begin{eqnarray}
H &=&H_{l}+H_{\text{sub}},  \notag \\
H_{l} &=&-J\overset{0}{\sum_{l=-\infty }}(a_{l}^{\dag }a_{l+1}+\text{H.c.}),
\\
H_{\text{sub}} &=&\sum_{i,j=1}^{N_{s}}\kappa _{ij}a_{i}^{\dag }a_{j}.  \notag
\end{eqnarray}%
It is noted that $H_{\text{sub}}$\ is non-Hermitian, possessing the
complex-valued eigen energy, while $H_{l}$\ is Hermitian, having complete
spectrum $E=-2J\cos \left( k\right) $, $\left( k\in \left[ 0,2\pi \right)
\right) $ and the eigen state $\sin \left( kj\right) a_{j}^{\dag }\left\vert
0\right\rangle $.\ To investigate the role of the lead in the non-Hermitian $%
H_{\text{sub}}$, we will consider the whole solution of the Hamiltonian $H$
and analyze its properties.

The eigen state can be expressed as $\left\vert \psi \right\rangle
=\sum_{j=-\infty }^{N_{s}}f^{k}\left( j\right) a_{j}^{\dag }\left\vert
0\right\rangle $. The explicit form of the wave function $f^{k}\left(
j\right) $ depends on the structure of $H_{\text{sub}}$. Generally speaking,
the solution of $f^{k}\left( j\right) $ can not be obtained exactly even the
explicit form of $H_{\text{sub}}$\ is given. However, within the lead the
wave function is always in the form
\begin{equation}
f^{k}\left( j\leq 0\right) =A_{k}e^{ikj}+B_{k}e^{-ikj},  \label{BA}
\end{equation}%
due to the semi-infinite boundary condition.

The Schr\"{o}dinger equation has the explicit form%
\begin{eqnarray}
&-Jf^{k}\left( j-1\right) -Jf^{k}\left( j+1\right) =Ef^{k}\left( j\right) ,%
\text{ }\left( j\leq 0\right) &  \notag \\
&-Jf^{k}\left( 0\right) +\sum_{i=1}^{N_{s}}\kappa _{i1}f^{k}\left( i\right)
=Ef^{k}\left( 1\right), &  \label{Seq} \\
&\sum_{i=1}^{N_{s}}\kappa _{ij}f^{k}\left( i\right) =Ef^{k}\left( j\right) ,%
\text{ }\left( j\in \left[ 2,N_{s}\right] \right) &  \notag
\end{eqnarray}%
within all the regions. The solutions of $E$ and $f^{k}\left( j\right) $
depend on the structure of the system $H_{\text{sub}}$. Nevertheless, the
exclusive geometry of the lead will give some clues to the characteristics
of the eigenvalues and eigenfunctions.

In the framework of Bethe ansatz method, all possible solutions within the
lead\ can be classified into three types:

(1) Scattering state wave, the wave function and energy are in the form
\begin{eqnarray}
f_{\text{\textrm{SS}}}\left( j\right) &=&A_{k}e^{ikj}+B_{k}e^{-ikj},\left(
k\ \text{is real}\right) \\
E &=&-2J\cos \left( k\right) .
\end{eqnarray}

(2) Monotonic damping wave, the wave function and energy are in the form%
\begin{eqnarray}
f_{\mathrm{MD}}\left( j\right) &=&\left( \pm 1\right) ^{j}e^{\beta j},\left(
\beta <0\right) \\
E &=&-2J\cosh \left( \beta \right) .
\end{eqnarray}

(3) Oscillation damping wave, the wave function and energy are in the form
\begin{eqnarray}
f_{\text{\textrm{OD}}}\left( j\right) &=&e^{ikj+\beta j},\left( \beta
<0,k\in \left( 0,2\pi \right) ,k\neq \pi \right) \\
E &=&-2J\left[ \cos \left( k\right) \cosh \left( \beta \right) +i\sin \left(
k\right) \sinh \left( \beta \right) \right] .
\end{eqnarray}%
%
%
%
%
%
%
%
%
%
%
%
%
%
%
%
%
%
%
%
%
%
%
%
%
%
%
%
%
%
%
%
%
%
%
%
%
%
%
%
%
%
%
%
%
%
%
%
%
%
%
%
%
%

\begin{figure}[tbp]
\includegraphics[ bb=89 450 547 620, width=0.4\textwidth, clip]{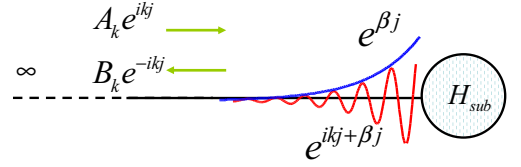}
\caption{(Color online) Schematic illustration of the configuration of the
concerned network and profiles of possible solutions. It consists of a
non-Hermitian finite cluster and a semi-infinite chain. The solutions on the
lead are threes types: scattering state wave (green), monotonic damping wave
(blue) and oscillation damping wave (red).}
\label{fig1}
\end{figure}

In Fig. \ref{fig1}, the concerned system and three types of possible
solutions within the lead\ are illustrated schematically. For the case of a
hermitian $H_{\text{sub}}$, the solutions are the form of $f_{\text{\textrm{%
SS}}}\left( j\right) $\ and $f_{\mathrm{MD}}\left( j\right) $\ with $%
\left\vert A\right\vert $ or $\left\vert B\right\vert =0$\ definitely. In
the case of non-Hermitian $H_{\text{sub}}$, $f_{\text{\textrm{OD}}}\left(
j\right) $ may appear associated with the complex energy level. In case of
absence of the solution $f_{\text{\textrm{OD}}}\left( j\right) $, full real
spectrum achieves, which shows the existence of the stationary states. It
indicates that the lead acts as a channel to balance the gain or loss in the
system $H_{\text{sub}}$.

At certain points $k_{c}$, the system makes transitions between
wavefunctions $f_{\text{\textrm{SS}}}\left( j\right) $ and $f_{\mathrm{MD}%
}\left( j\right) $,\ as well as between $f_{\text{\textrm{SS}}}\left(
j\right) $ and $f_{\text{\textrm{OD}}}\left( j\right) $. The former
transition is actually a switch between real and imaginary $k$, preserving
the reality of the eigen energy. Then the transition point locates at $%
k=0,\pi $,\ i.e.,%
\begin{eqnarray}
&&f_{\mathrm{MD}}\left( j\right) \overset{\beta =0}{\longrightarrow }\left(
\pm 1\right) ^{j}, \\
&&f_{\text{\textrm{SS}}}\left( j\right) \overset{A\text{ \textrm{or} }B=0,%
\text{ }k=0,\pi }{\longrightarrow }\left( \pm 1\right) ^{j},
\end{eqnarray}%
which usually occurs in the case of Hermitian $H_{\text{sub}}$. The later
transition only occurs in a non-Hermitian system, eigen energy switching
between real and complex values. In contrast to the above case, the
transition point (referred as exceptional point) depends on the structure of
the non-Hermitian $H_{\text{sub}}$, i.e.,
\begin{eqnarray}
&&f_{\text{\textrm{OD}}}\left( j\right) \overset{\beta =0}{\longrightarrow }%
e^{ik_{c}j}, \\
&&f_{\text{\textrm{SS}}}\left( j\right) \overset{B=0}{\longrightarrow }%
e^{ik_{c}j}.
\end{eqnarray}%
It indicates that a unidirectional plane wave exists in the lead when an
appropriate\ non-Hermitian $H_{\text{sub}}$ is connected. It has both
fundamental as well as practical implications. This result reveals the
exceptional point from an alternative way: It is the threshold of the
balance between the non-Hermitian\ subcluster and the lead. From a practical
perspective, the unidirectional-plane-wave solution at the exceptional point
can be used to realize the reflectionless absorption and persistent emission
in the experiment. To characterize the probability generation (negative in
the case of the dissipation) of the non-Hermitian cluster, we introduce the
current operator \cite{Caroli}%
\begin{equation}
\mathcal{\hat{J}}_{j}=-iJ\left( a_{j}^{\dagger }a_{j+1}-a_{j+1}^{\dagger
}a_{j}\right) ,
\end{equation}%
where $j\in \left[ -\infty ,0\right] $.

For three types of wave functions $f_{\text{\textrm{SS}}}\left( j\right)
e^{-iEt}$, $f_{\mathrm{MD}}\left( j\right) e^{-iEt}$\ and $f_{\text{\textrm{%
OD}}}\left( j\right) e^{-iEt}$, the corresponding currents can be obtained
as
\begin{eqnarray}
\mathcal{J}_{\text{\textrm{SS}}} &=&2J(\left\vert A_{k}\right\vert
^{2}-\left\vert B_{k}\right\vert ^{2})\sin \left( k\right) ,  \notag \\
\mathcal{J}_{\text{\textrm{MD}}} &=&0, \\
\mathcal{J}_{\text{\textrm{OD}}} &=&2J\sin \left( k\right) e^{2\left[ \beta
j-2J\sin \left( k\right) \sin \left( \beta \right) t\right] }.  \notag
\end{eqnarray}%
We can see that $\mathcal{J}_{\text{\textrm{SS}}}$\ is time-independent and
is conservative along the lead, representing a steady flow or the dynamic
balance, while $\mathcal{J}_{\text{\textrm{OD}}}$\ is non-periodically
time-dependent, indicating the unbalance of the state. In other word, the
mechanism of the reality of the spectrum is the balance between the source
(or drain) and the channel of the probability flow. Then the exceptional
point is the threshold of such dynamic balance, corresponding to the
unidirectional-plane-wave, i.e., $\left\vert A_{k}\right\vert =0$\ or $%
\left\vert B_{k}\right\vert =0$. Then the probability generation for the
exceptional point is%
\begin{equation}
\mathcal{J}_{c}=2J\sin \left( k_{c}\right) ,  \label{J_c}
\end{equation}%
which the sign indicates that the cluster is a source or drain, then it is
referred as critical current in this paper. Unlike the situation in
traditional quantum mechanics, the magnitude of the current $\mathcal{J}_{c}$%
\ does not represent the absolute current in traditional quantum mechanics
because the corresponding eigenstate is not normalized under the Dirac inner
product.

We will demonstrate and explain these points through the following
illustrative example. We would like to point out that, there is another type
of the exceptional point, arising from the transition of two types of wave
functions $f_{\text{\textrm{OD}}}\left( j\right) $ and $f_{\text{\textrm{MD}}%
}\left( j\right) $, which is beyond our interest.

\subsection{Illustrative examples}

\label{illustrative examples}In this subsection, we investigate two simple
exactly solvable systems to illustrate the main idea of this paper. In order
to exemplify the above mentioned analysis of relating the wavefunction
within the lead and the eigenvalue, we take $H_{\text{sub}}$\ to be the
simplest non-Hermitian networks to construct two types of exemplified
systems. Type I is a uniform chain with a complex potential at one end and
type II is a uniform chain with a complex hopping at one end. In the
following, we present the analytical results in the framework of above
mentioned for the two models in order to perform a comprehensive study.


\begin{figure}[tbp]
\includegraphics[bb=48 75 548 726, width=0.4\textwidth, clip]{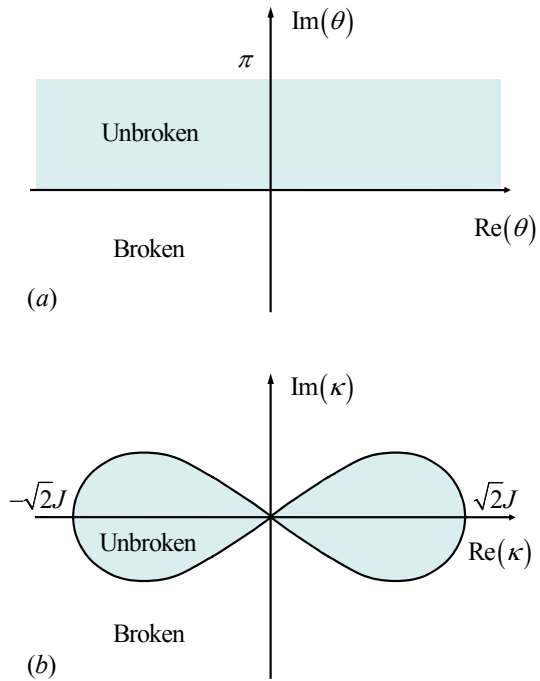}
\caption{(Color online) Schematic illustration of the phase diagrams in the
infinite N system: (a) a uniform chain with a complex potential and (b) with
a complex hopping at one end. }
\label{fig2}
\end{figure}


\subsubsection{Semi-infinite system with complex potential}

\label{eg_complex_potential}
\begin{figure}[tbp]
\includegraphics[bb=27 179 542 685, width=0.4\textwidth, clip]{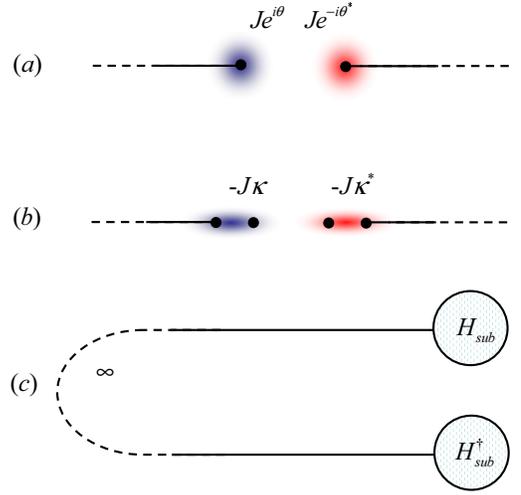}
\caption{(Color online) Schematic illustration for the non-Hermitian $%
\mathcal{PT}$-symmetric networks: (a) two separable semi-infinite chains
with the complex potential and (b) with the complex hopping at the end. (c)
When an arbitrary semi-infinite system and its conjugate counterpart are
connected at infinity, a $\mathcal{PT}$-symmetric system are constructed.}
\label{fig3}
\end{figure}

The type I Hamiltonian has the form
\begin{equation}
H_{\text{CP}}=-J\sum_{j=1}^{\infty }(a_{j}^{\dagger }a_{j+1}+\text{H.c.}%
)+Je^{i\theta }a_{1}^{\dagger }a_{1},  \label{H_CP}
\end{equation}%
where $\theta $\ is a complex number. According to Bethe ansatz method, the
wavefunction $f^{k}\left( j\right) $\ can be expressed as

\begin{equation}
f^{k}\left( j\right) =A_{k}e^{ikj}+B_{k}e^{-ikj},j\in \left[ 0,\infty \right)
\end{equation}%
and the Schr\"{o}dinger equations for $H_{\text{CP}}$ is

\begin{eqnarray}
-Jf^{k}\left( j+1\right) -Jf^{k}\left( j-1\right)  &=&E_{k}f^{k}\left(
j\right) , \\
-Jf^{k}\left( 2\right)  &=&\left[ E_{k}-Je^{i\theta }\right] f^{k}\left(
1\right) .  \notag
\end{eqnarray}%
Submitting $f^{k}\left( j\right) $\ into the Schr\"{o}dinger equation, we
have%
\begin{equation}
R_{k}=\frac{B_{k}}{A_{k}}=-\frac{1+e^{i\left( \theta +k\right) }}{%
1+e^{i\left( \theta -k\right) }},  \label{R_cp}
\end{equation}%
which is the reflection amplitude for the scattering state. Now we are
interested in the wavefunction with complex eigen energy. The existence of
the solution $f_{\text{\textrm{OD}}}\left( j\right) $\ requires%
\begin{equation}
e^{-ik}=-e^{i\theta }\text{ and Im}\left( k\right) >0,
\end{equation}%
which lead to $k=\pi -\theta $ with Im$\left( \theta \right) <0$. Then we
conclude that there is a unique complex solution within the region Im$\left(
\theta \right) <0$ and the system has full real spectrum if the potential is
in the rest region. At the boundary, we have
\begin{equation}
\text{Re}\left( \theta _{c}\right) =\theta _{c},
\end{equation}%
which indicates a circle of radius $J$ in the complex plane.\ The phase
diagram is sketched in Fig. \ref{fig3} (a). Then the corresponding
wavefunction has the form%
\begin{equation}
f^{k_{c}}\left( j\right) =e^{i\left( \pi -\theta _{c}\right) j},
\label{potentialk}
\end{equation}%
which represents a unidirectional plane wave with energy%
\begin{equation}
E_{k_{c}}=2J\cos \left( \theta _{c}\right) .
\end{equation}%
Accordingly, the critical current is
\begin{equation}
\mathcal{J}_{c}=2J\sin \left( \theta _{c}\right) ,
\end{equation}%
which accords with the intuition that a positive imaginary potential can be
a source and a negative imaginary potential can be a drain. However, unlike
the situation in traditional quantum mechanics, the magnitude of the current
$\mathcal{J}_{c}$\ does not represent the absolute current in traditional
quantum mechanics because the eigenstate is not normalized under the Dirac
inner product.


Before further discussion of the implication of the obtained result, two
distinguishing features need to be mentioned. Firstly, the non-Hermitian
system can have full real spectrum even though there is no symmetry
required. Secondly, there is only one possible complex energy level rather
than complex conjugate pairs. Thus there is no level coalescing occurring at
the exceptional point. Both of these differ from that of a finite $\mathcal{%
PT}$-symmetric system. In the following example, it will be shown that such
features are not exclusive to the complex potential.

\subsubsection{Semi-infinite system with complex-coupling dimer}

\label{eg_Complex-coupling}The type II Hamiltonian has the form%
\begin{equation}
H_{\text{CC}}=-J\sum_{j=2}^{\infty }(a_{j}^{\dagger }a_{j+1}+\text{H.c.}%
)-J\kappa a_{1}^{\dagger }a_{2}-J\kappa a_{2}^{\dagger }a_{1},  \label{Hcc}
\end{equation}%
which is a different type of the non-Hermitian model in contrast to the type
I. Here $\kappa $\ is a complex number. In the following, we will perform a
parallel investigation with the current Hamiltonian.\ The Bethe ansatz wave
function has the form%
\begin{equation}
f^{k}\left( j\right) =\left\{
\begin{array}{cc}
A_{k}e^{ikj}+B_{k}e^{-ikj}, & j\in \left[ 1,\infty \right)  \\
C_{k}e^{ikj}+D_{k}e^{-ikj}, & j=0%
\end{array}%
\right. .
\end{equation}%
Substituting $f^{k}\left( j\right) $\ to the Schr\"{o}dinger equation,%
\begin{eqnarray}
-Jf^{k}\left( j-1\right) -Jf^{k}\left( j+1\right)  &=&E_{k}f^{k}\left(
j\right) ,j\in \left[ 3,\infty \right)   \notag \\
-J\kappa f^{k}\left( 1\right) -Jf^{k}\left( 3\right)  &=&E_{k}f^{k}\left(
2\right) , \\
-J\kappa f^{k}\left( 2\right)  &=&E_{k}f^{k}\left( 1\right) ,  \notag
\end{eqnarray}%
we obtain the reflection amplitude%
\begin{equation}
R_{k}=\frac{B_{k}}{A_{k}}=-\frac{\left( \kappa ^{2}-1\right) e^{2ik}-1}{%
\left( \kappa ^{2}-1\right) e^{-2ik}-1}.  \label{R_cc}
\end{equation}%
The existence of the solution $f_{\text{\textrm{OD}}}\left( j\right) $\
requires%
\begin{equation}
k=\frac{1}{2}i\ln \left( \kappa ^{2}-1\right) \text{, }\frac{1}{2}i\ln
\left( \kappa ^{2}-1\right) +\pi \text{ and Im}\left( k\right) >0.
\end{equation}%
Similarly, we conclude that there are two complex solutions within the
region $\left\vert \kappa \right\vert ^{4}-2$Re$\left( \kappa ^{2}\right) >0$%
, and the system has full real spectrum if $\kappa $\ is in the region $%
\left\vert \kappa \right\vert ^{4}-2$Re$\left( \kappa ^{2}\right) \leqslant 0
$. The phase diagram is sketched in Fig. \ref{fig2} (b). The boundary can be
expressed as%
\begin{equation}
\frac{1}{2}\left\{ \left[ \text{Re}\left( \kappa _{c}\right) \right] ^{2}+%
\left[ \text{Im}\left( \kappa _{c}\right) \right] ^{2}\right\} ^{2}=\left[
\text{Re}\left( \kappa _{c}\right) \right] ^{2}-\left[ \text{Im}\left(
\kappa _{c}\right) \right] ^{2},  \label{boundary_cc}
\end{equation}%
which is a Lemniscate of Bernoulli in the complex plane. Then the
corresponding eigen wave functions at the boundary have the form%
\begin{equation}
f^{k_{c}}\left( j\right) =\left\{
\begin{array}{cc}
\pm e^{i\left( \varphi /2\right) j}, & j\in \left[ 2,\infty \right)  \\
\pm \kappa _{c}^{-1}e^{i\left( \varphi /2\right) j}, & j=1%
\end{array}%
\right. ,
\end{equation}%
where
\begin{equation}
\tan \varphi =i\frac{\kappa _{c}^{2}-\left( \kappa _{c}^{\ast }\right) ^{2}}{%
\left\vert \kappa _{c}\right\vert ^{4}-2}  \label{cc1}
\end{equation}%
is real. The wave functions $f^{k_{c}}\left( j\right) $\ represent
unidirectional plane waves with energy $E_{k_{c}}=\pm 2J\cos \left( \varphi
/2\right) $, respectively. This indicates that a complex-coupling dimer is a
different type of basic element for a discrete non-Hermitian system in
comparison with complex potential. There are two eigenstates corresponding
to a single exceptional point. In virtue of the critical current $J_{c}=\pm
2J\sin \left( \varphi /2\right) $, one can see that the complex-coupling
dimer\ acts as a source for $J_{c}>0$\ but a drain for $J_{c}<0$.

\begin{figure}[tbp]
\includegraphics[ bb=120 300 480 631, width=0.3\textwidth, clip]{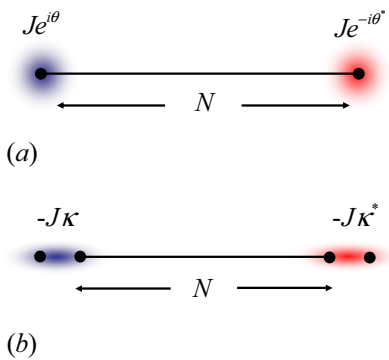}
\caption{(Color online) Schematic illustration for the non-Hermitian $%
\mathcal{PT}$-symmetric networks: (a) an uniform chain with a complex
potential and (b) with a complex hopping at both sides.}
\label{fig4}
\end{figure}

It is noted that both above two examples are not symmetric, which show that
the symmetry is not the necessary condition for the occurrence of full real
spectrum. The underlying mechanism can be explained as the balance between
the source (or drain) and the channel. In this sense, a semi-infinite chain
can act as a source (or drain) to balance the original drain (or source).
The exceptional point is the threshold of such balance. This point will be
elucidated in details in the section \ref{relation with PT}.

\begin{figure*}[tbp]
\includegraphics[ bb=16 177 546 595, width=6.5 cm, clip]{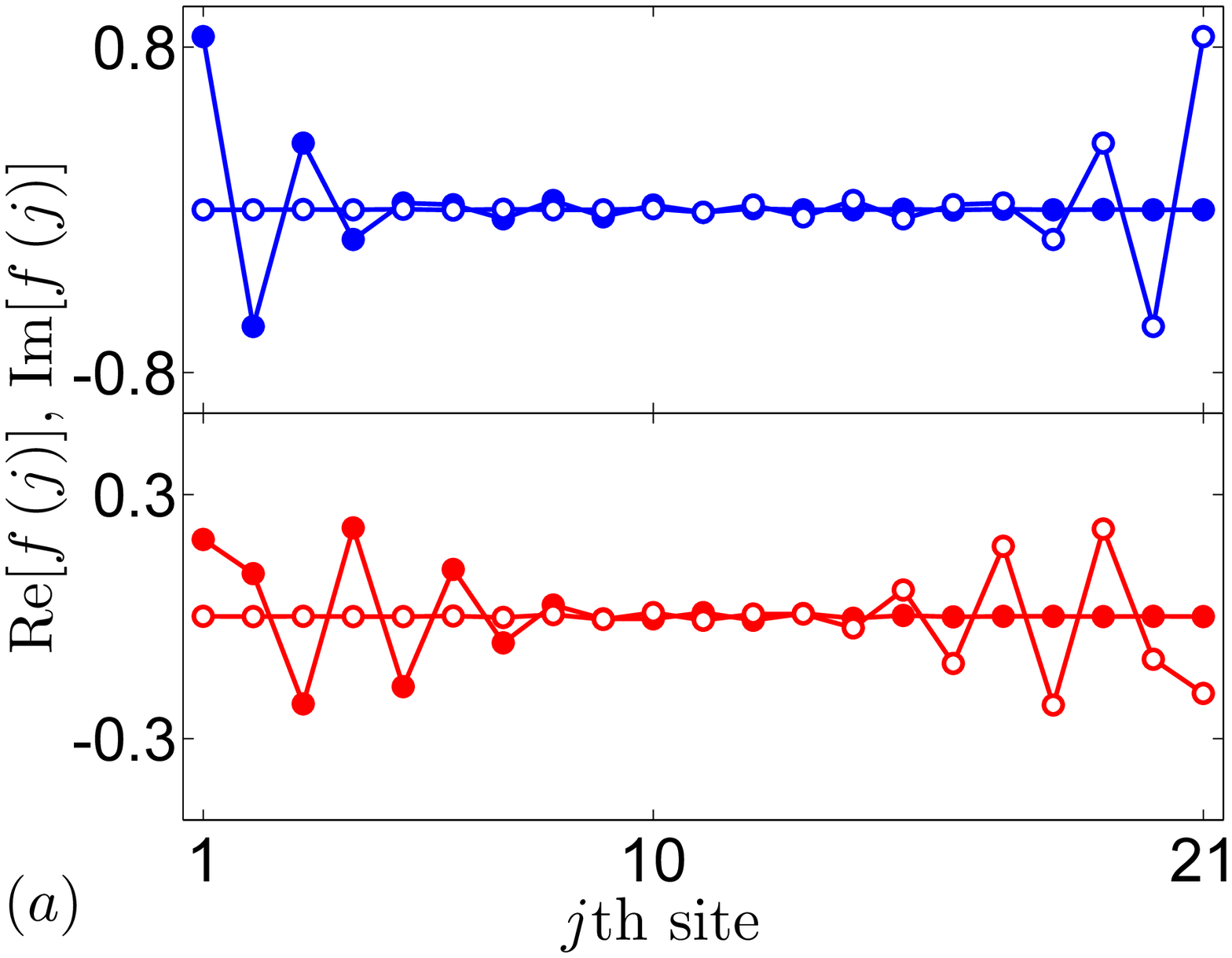} %
\includegraphics[ bb=16 177 546 595, width=6.5 cm, clip]{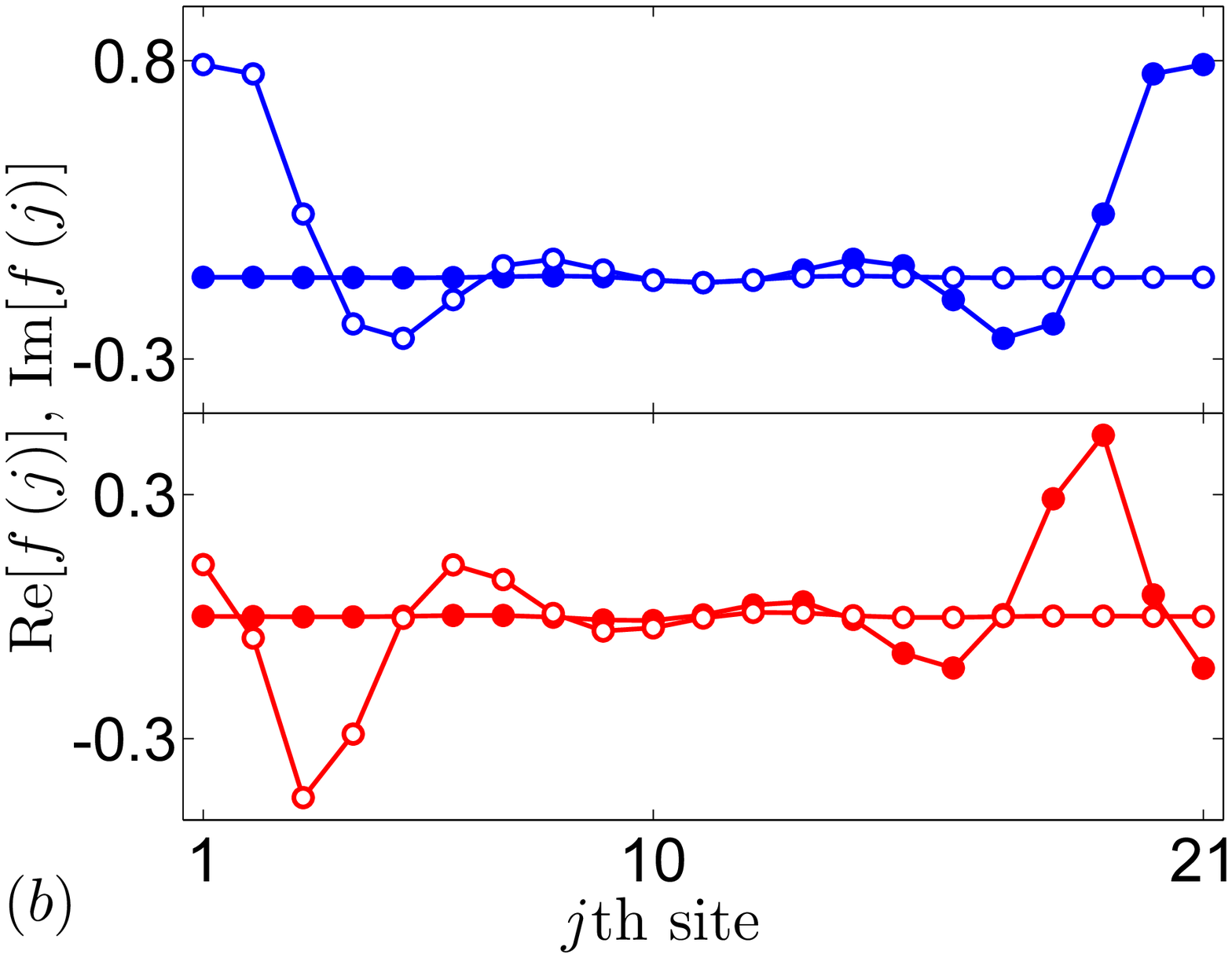}
\caption{(Color online) Plots of the eigen fuctions of (a) $H_{\text{1}}$\
with $\protect\theta =0.4-0.4$i, eigen energy $1.99\pm 0.32$i, and (b) $H_{%
\text{2}}$\ with $\protect\kappa =1+1$i, eigen energy $-1.14\pm 0.71$i, for $%
N=20$\ chain. The blue and red solid (empty) circle indicate the real
(imaginary) part of wave functions for the conjugate pairs, respectively. It
can be observed that the eigen functions are localized around the
non-Hermitian clusters and thus break the $\mathcal{PT}$\ symmetry. We also
note that two eigen functions of the conjugate pairs are $\mathcal{PT}$%
-symmetric counterpart to each other.}
\label{fig5}
\end{figure*}


\section{Connection between the semi-infinite systems and $\mathcal{PT}$
symmetric systems}

\label{relation with PT}So far we have shown that there exist a class of
semi-infinite non-Hermitian non-$\mathcal{PT}$\ symmetric systems possessing
fully real spectra. The occurrence of complex levels in such systems, or
quantum transition, is not accompanied by the spontaneous symmetry breaking,
but the delocalization-localization transition of wave functions. So it is
interesting to consider the connection between the obtained results and the
well developed non-Hermitian $\mathcal{PT}$-symmetric quantum mechanics.

\subsection{$\mathcal{PT}$ symmetric infinite systems}

We start the analysis with the simple configuration, which can constructed
from $H_{\text{CP}}$ ($H_{\text{CC}}$) and its $\mathcal{PT}$\ counterpart $%
\mathcal{PT}H_{\text{CP}}\left( \mathcal{PT}\right) ^{-1}$ ($\mathcal{PT}H_{%
\text{CC}}\left( \mathcal{PT}\right) ^{-1}$), as sketched in Figs. \ref{fig3}
(a) and (b). Here the action of the parity operator $\mathcal{P}$ is defined
as $\mathcal{P}:l\rightarrow -l$ and the time-reversal operator $\mathcal{T}$
as $\mathcal{T}:i\rightarrow -i$.\ The Hamiltonians are written as%
\begin{eqnarray}
H_{\text{CP}}^{\mathcal{PT}} &=&H_{\text{CP}}+\mathcal{PT}H_{\text{CP}%
}\left( \mathcal{PT}\right) ^{-1}  \label{H_PTCP} \\
&=&-J\sum_{\substack{ j=-\infty , \\ \left( j\neq -1,0\right) }}^{\infty
}(a_{j}^{\dagger }a_{j+1}+\text{H.c.})  \notag \\
&&+Je^{i\theta }a_{1}^{\dagger }a_{1}+Je^{-i\theta ^{\ast }}a_{-1}^{\dagger
}a_{-1}  \notag
\end{eqnarray}%
and%
\begin{eqnarray}
H_{\text{CC}}^{\mathcal{PT}}=H_{\text{CC}}+\mathcal{PT}H_{\text{CC}}\left(
\mathcal{PT}\right) ^{-1} &&  \label{H_PTCC} \\
=-J\sum_{\substack{ j=-\infty , \\ \left( j\neq -2,\pm 1,0\right) }}^{\infty
}(a_{j}^{\dagger }a_{j+1}+\text{H.c.}) &&  \notag \\
-J\kappa \left( a_{1}^{\dagger }a_{2}+\text{H.c.}\right) -J\kappa ^{\ast
}\left( a_{-1}^{\dagger }a_{-2}+\text{H.c.}\right) . &&  \notag
\end{eqnarray}%
Obviously, they are $\mathcal{PT}$-symmetric and have all features of
typical pseudo-Hermitian systems: (i) They have the same phase diagrams in
Fig. (\ref{fig2}) as their corresponding sub-systems $H_{\text{CP}}$\ and $%
H_{\text{CC}}$. At the exceptional points, two eigen functions coalesce an
entire plane wave within the whole space except the $0$th site. (ii) The
complex levels come in conjugate pairs. $\mathcal{PT}$-symmetry of the eigen
functions break. This toy model shows us the essence of symmetry breaking:
the delocalization-localization transition of wave functions in each
sub-system. At this point we are ready to move to a more complete
description, considering an inseparable system.

We start from the corresponding solutions of the Hamiltonians $H_{\text{CP}%
}^{\dag }$\ and $H_{\text{CC}}^{\dag }$, which actually can be obtained by
applying time-reversal operation, i.e., taking complex conjugation for the
obtained solutions of the Hamiltonians $H_{\text{CP}}$\ and $H_{\text{CC}}$.
It is interesting to find that the real eigen-valued solutions between $H_{%
\text{CP}}$\ and $H_{\text{CP}}^{\dag }$\ ($H_{\text{CC}}$\ and $H_{\text{CC}%
}^{\dag }$) can match with each other due to the fact that%
\begin{equation}
\frac{A_{k}}{B_{k}}=\frac{B_{-k}}{A_{-k}},
\end{equation}%
for both examples in the Eqs. (\ref{R_cp}) and (\ref{R_cc}). In other words,
all the real eigen-valued solutions will not change if two systems $H_{\text{%
CP}}$\ and $H_{\text{CP}}^{\dag }$\ ($H_{\text{CC}}$\ and $H_{\text{CC}%
}^{\dag }$) are connected at the infinity, as sketched in Fig. \ref{fig3}
(c). It is presumable that the similar situation could occur in finite
system, i.e., a semi-infinite non-$\mathcal{PT}$ symmetric system can be
regarded as the rudiment of the corresponding finite $\mathcal{PT}$%
-symmetric system. We will demonstrate this point through the following
illustrative examples, which are finite versions of combined $H_{\text{CP}}$%
\ and $H_{\text{CP}}^{\dag }$\ ($H_{\text{CC}}$\ and $H_{\text{CC}}^{\dag }$%
). A sketch of such systems are given in Figs. \ref{fig4} (a) and \ref{fig4}
(b).

\subsection{$\mathcal{PT}$ symmetric finite systems}

The potential example\textbf{\ }is a $\mathcal{PT}$ symmetric\ non-Hermitian
$N$-site chain with complex on-site potential at two ends, which has the
Hamiltonian
\begin{equation}
H_{\text{1}}=-J\sum_{j=1}^{N-1}(a_{j}^{\dag }a_{j+1}+\text{H.c.}%
)+Je^{i\theta }a_{1}^{\dag }a_{1}+Je^{-i\theta ^{\ast }}a_{N}^{\dag }a_{N}%
\text{.}  \label{H_1}
\end{equation}%
It is a $\mathcal{PT}$ symmetric model, i.e., $[\mathcal{PT},H_{\text{1}}]=0$%
, where the action of the parity operator $\mathcal{P}$ is defined as $%
\mathcal{P}:l\rightarrow N+1-l$ and the time-reversal operator $\mathcal{T}$
as $\mathcal{T}:i\rightarrow -i$.\ For infinite $N$, it becomes the
combination of the systems $H_{\text{CP}}$\ and $H_{\text{CP}}^{\dag }$. For
finite $N$, it is an extension version of the model proposed in the previous
paper Ref. \cite{Liang Jin}. By using the standard Bethe ansatz method, the
solution is determined by the critical equation
\begin{equation}
\Gamma \left( k\right) =0  \label{tau_k}
\end{equation}%
where%
\begin{eqnarray}
\Gamma \left( k\right)  &=&e^{i\left( \theta -\theta ^{\ast }\right) }\sin
\left[ k(N-1)\right]  \\
&&+\sin \left[ k(N+1)\right] +2\text{Re}\left( e^{i\theta }\right) \sin
\left( kN\right) .  \notag
\end{eqnarray}%
Accordingly, the exceptional point can be obtained by the equations \cite%
{Liang Jin,L. Jin}%
\begin{equation}
\Gamma \left( k_{c}\right) =0\text{ and }\left. \frac{\text{d}\Gamma \left(
k\right) }{\text{d}k}\right\vert _{k=k_{c}}=0.  \label{Critical_condition}
\end{equation}%
It is difficult to get the explicit solutions of the Eq. (\ref%
{Critical_condition}) for finite $N$. Nevertheless, the equation about d$%
\Gamma \left( k\right) /$d$k$\ can be reduced to%
\begin{eqnarray}
e^{i\left( \theta -\theta ^{\ast }\right) }\cos \left[ k_{c}(N-1)\right]
+\cos \left[ k_{c}(N+1)\right]  && \\
+2\text{Re}\left( e^{i\theta }\right) \cos \left( k_{c}N\right) \approx 0 &&
\notag
\end{eqnarray}%
by taking the approximation $N\pm 1\approx N$ in the large $N$\ limit. It is
easy to find that the existence of real $k_{c}$ solution\ requires Im$\left(
\theta \right) =0$ and $k_{c}=\pi -\theta $. It is in accordance with result
in the corresponding semi-infinite system.
\begin{figure*}[tbp]
\includegraphics[ bb=22 166 570 618, width=6.5 cm, clip]{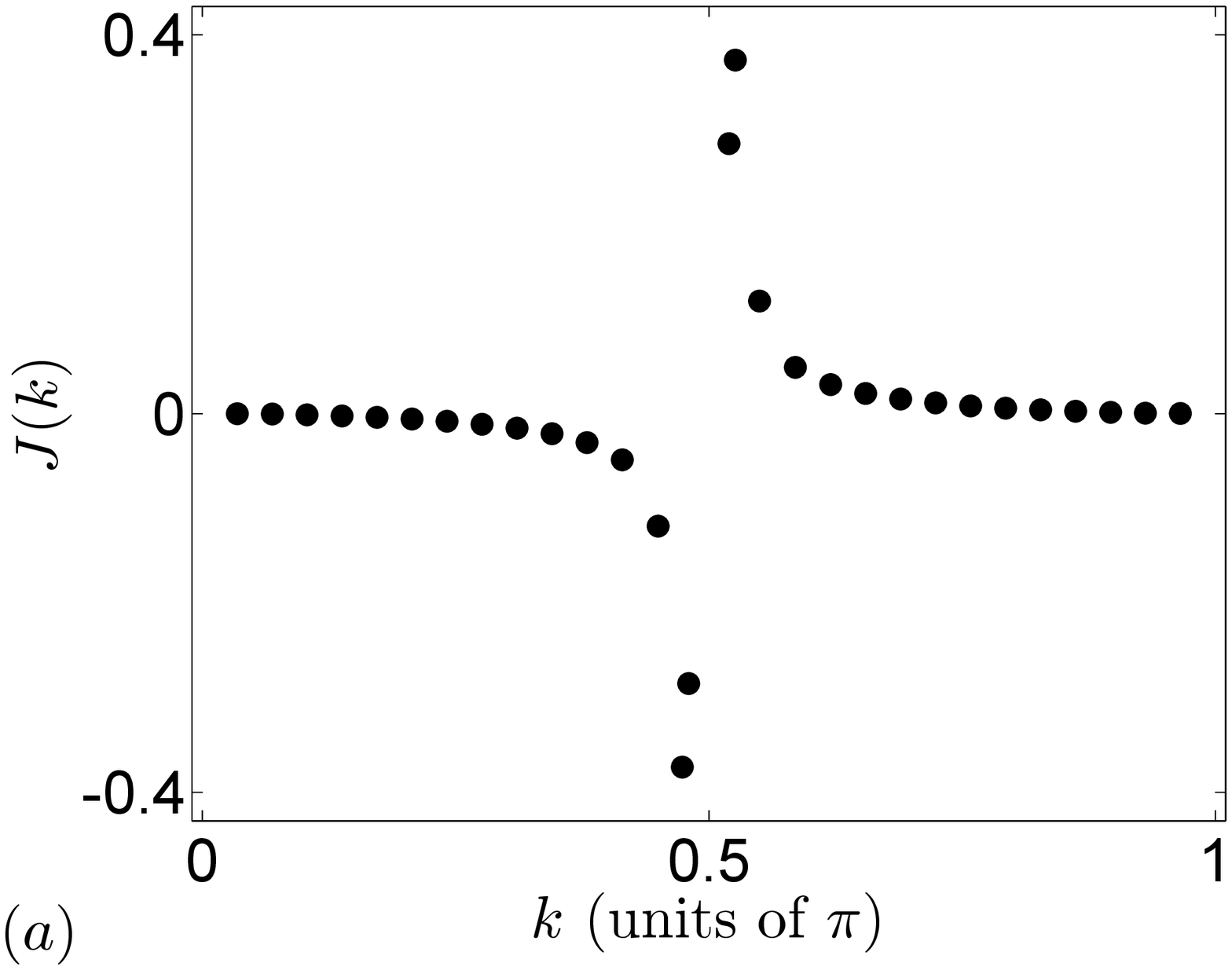} %
\includegraphics[ bb=22 166 570 618, width=6.5 cm, clip]{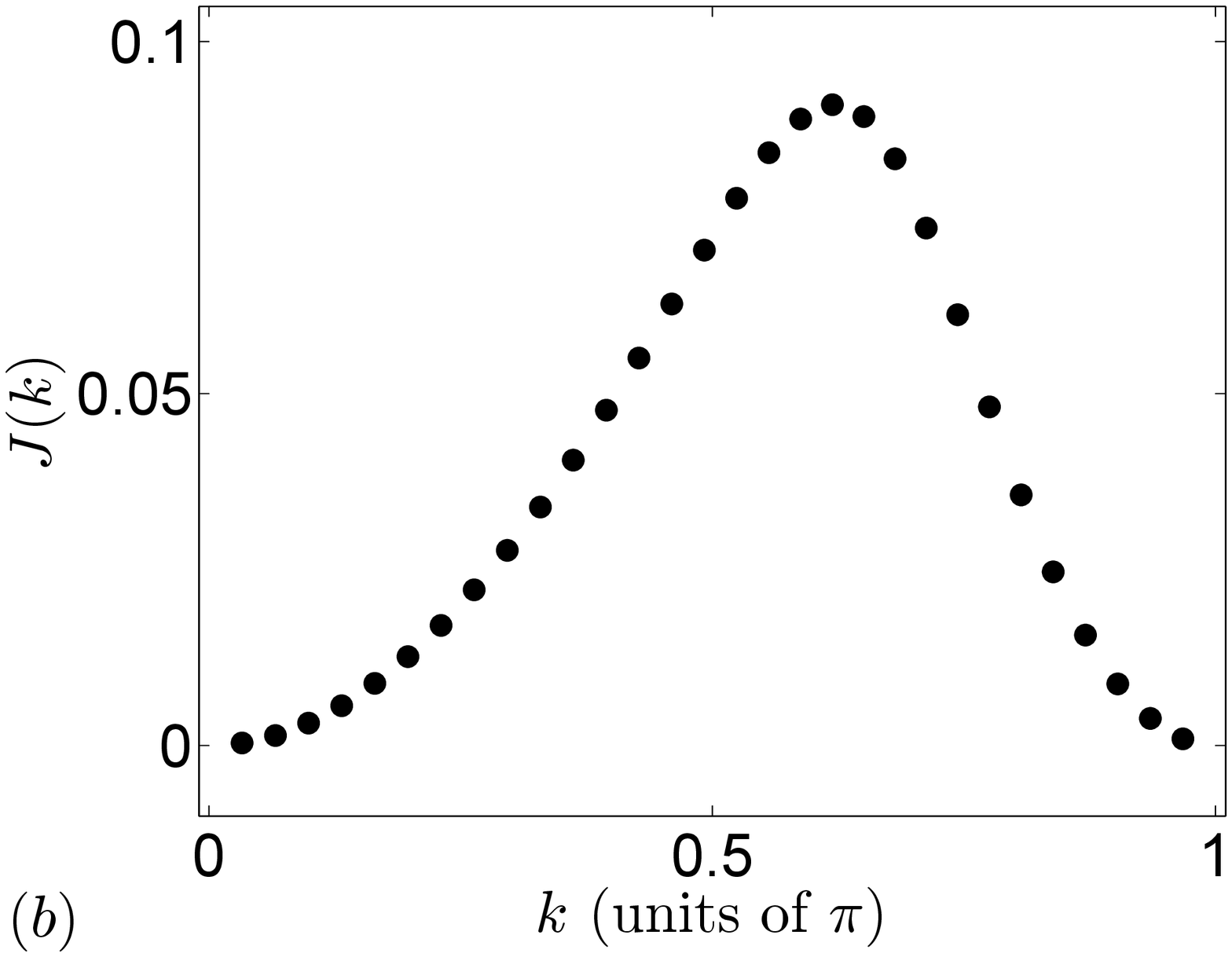}
\caption{(Color online) Plots of the current $\mathcal{J}\left( k\right) $\
for eigenstates of (a) $H_{\text{1}}$\ with $\protect\theta =1+0.5$i and (b)
$H_{\text{2}}$\ with $\protect\kappa =0.3+0.3$i, for $N=30$\ chain. As
mentioned in the text, the magnitude itself of the current is meaningless.
We are interested in the sign of it, which indicates direction of flow, or
the location of the source. We note that the sign of the current for $H_{%
\text{1}}$\ are the same and the complex potential can always be source when
its imaginary part is positive. This is in accordance with our intuition.
Nevertheless, the sign of the current for $H_{\text{2}}$\ are different,
which indicates that a non-Hermitian dimer cannot be a source or drain
absolutely.}
\label{fig6}
\end{figure*}

The dimer example\textbf{\ }can be described by the Hamiltonian%
\begin{eqnarray}
H_{\text{2}} &=&-J\sum_{j=2}^{N-2}(a_{j}^{\dag }a_{j+1}+\text{\textrm{H.c.}}%
)-J\kappa (a_{1}^{\dag }a_{2}+\text{\textrm{H.c.}})  \notag \\
&&+J\kappa ^{\ast }\left( a_{N-1}^{\dag }a_{N}+\text{\textrm{H.c.}}\right) ,
\label{H_2}
\end{eqnarray}%
which corresponds to the combination of two systems $H_{\text{CC}}$\ and $H_{%
\text{CC}}^{\dag }$. By the same procedure as that for $H_{\text{1}}$, we
find that the critical equation for finite $N$ is
\begin{eqnarray}
\chi _{2}\sin \left[ k\left( N-3\right) \right] +\chi _{1}\sin \left[
k\left( N-1\right) \right]  && \\
-\sin \left[ k\left( N+1\right) \right] =0 &&  \notag
\end{eqnarray}%
where%
\begin{eqnarray}
\chi _{1} &=&\kappa ^{2}+\left( \kappa ^{\ast }\right) ^{2}-2, \\
\chi _{2} &=&\chi _{1}-\left\vert \kappa \right\vert ^{4}+1.
\end{eqnarray}%
In addition, the exceptional point for large $N$ is determined by the
equations
\begin{equation}
2\chi _{2}\cos \left( 2k_{c}\right) +\chi _{1}\approx 0
\end{equation}%
and%
\begin{equation}
2\cos \left( 2k_{c}\right) -\chi _{1}=0,
\end{equation}%
which leads to the same results as the Eqs. (\ref{boundary_cc}) and (\ref%
{cc1}).

It has been shown by the\textbf{\ }$\mathcal{PT}$-symmetric quantum theory,
beyond the exceptional points the complex levels in both above two models
come in conjugate pairs \cite{A.M43} and the $\mathcal{PT}$\ symmetry of%
\textbf{\ }the corresponding eigen functions break. Nevertheless according
to our above analysis, the occurrence of the complex level should be
accompanied by the delocalization-localization transition of the
corresponding wave function. We perform numerical simulation of eigen
functions for finite size system to demonstrate this connection. In Fig. \ref%
{fig5}, we plot the eigenfuctions with complex eigen values including real
and imaginary parts for the models $H_{\text{1}}$\ and $H_{\text{2}}$,
respectively.\textbf{\ }It shows that the\textbf{\ }$\mathcal{PT}$\textbf{\ }%
symmetry\ of all the eigen functions break and are local, which accords with
our analysis.
\begin{figure*}[tbp]
\includegraphics[ bb=24 174 540 606, width=6.5 cm, clip]{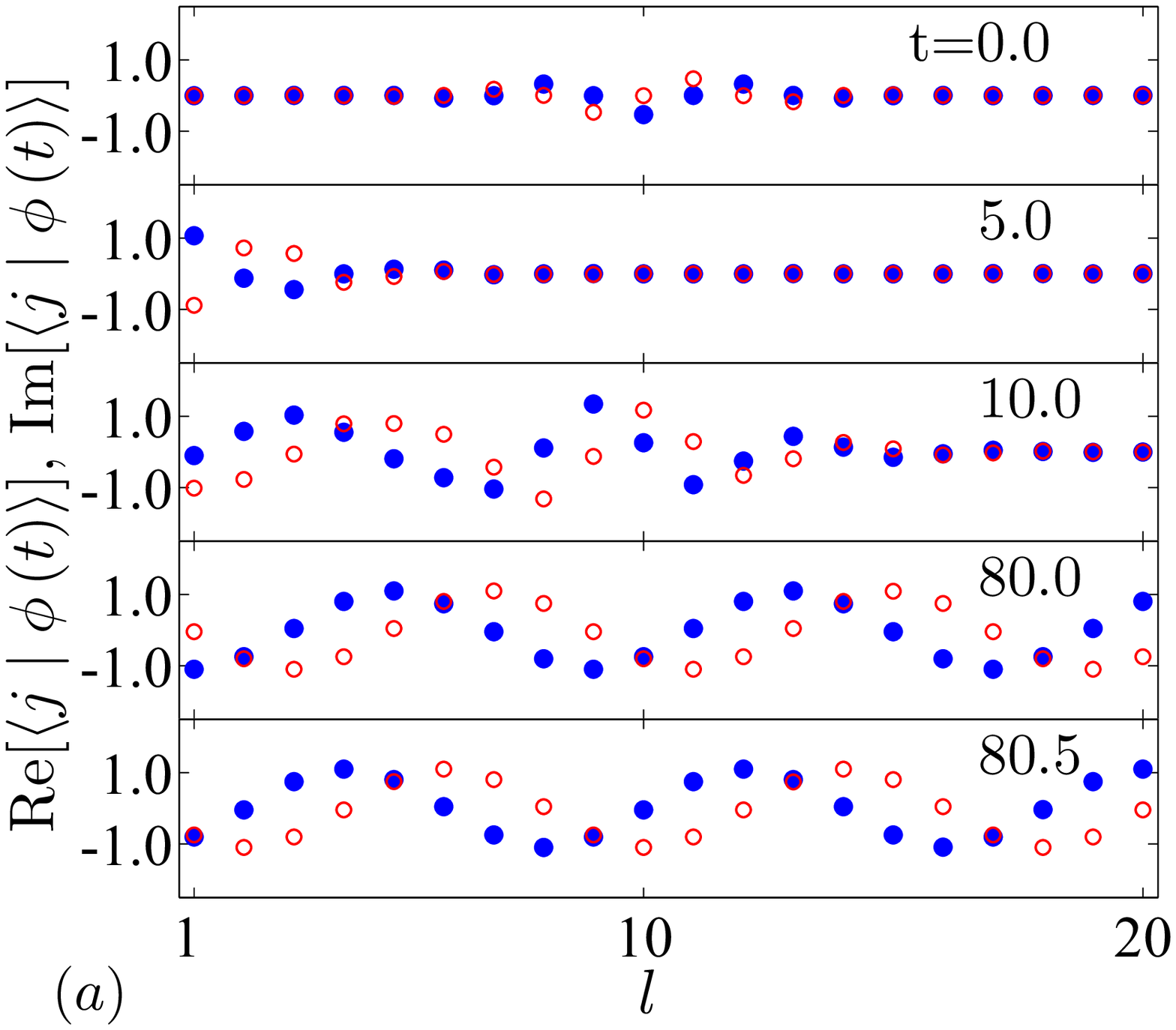} %
\includegraphics[ bb=2 174 518 606, width=6.5 cm, clip]{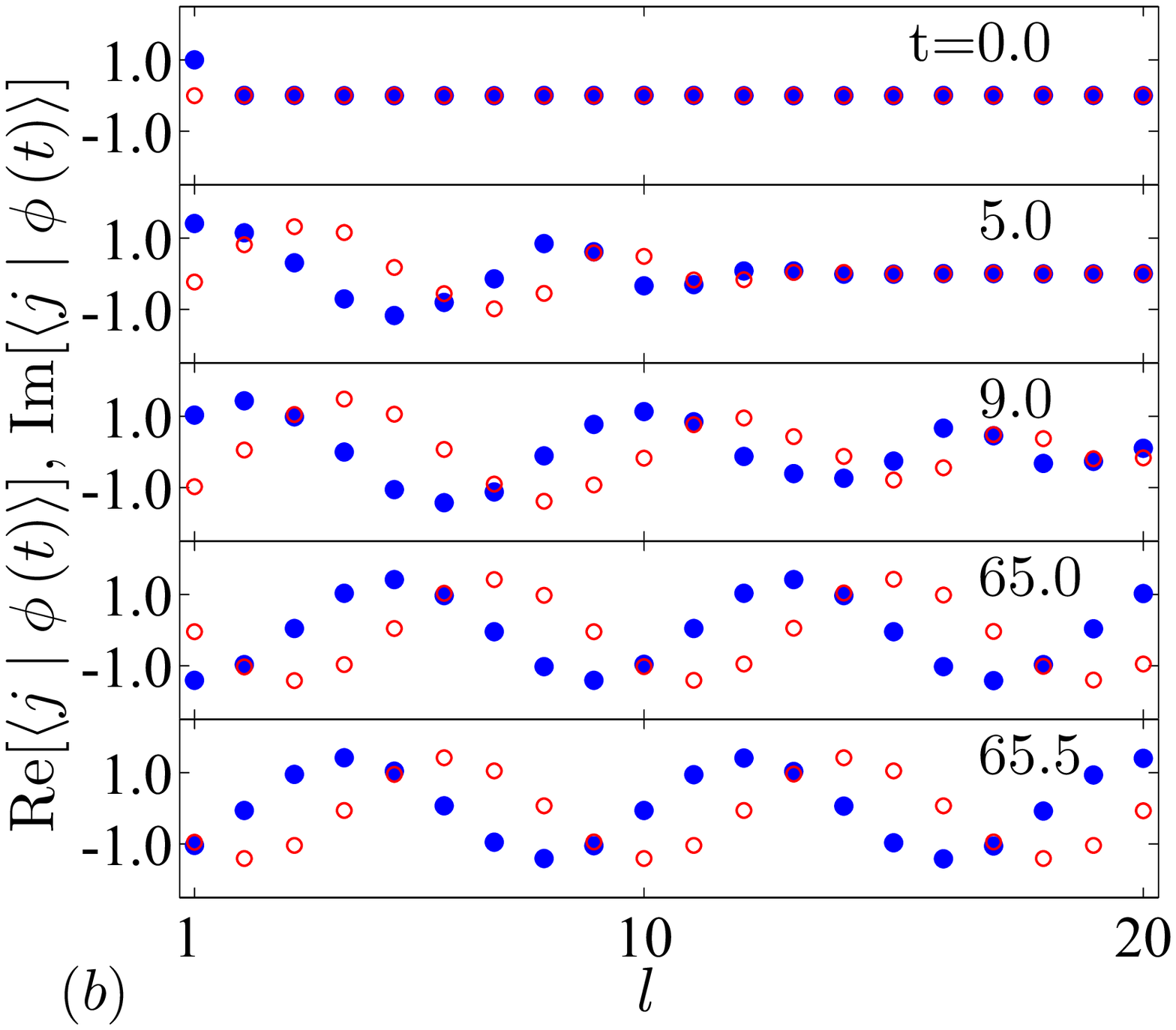}
\caption{(Color online) The profile of the evolved wave function $%
\left\langle j\right. \left\vert \protect\phi \left( t\right) \right\rangle $
for the initial state being (a) an incident Gaussian wavepacket with $k_{0}=-%
\protect\pi /2$, $N_{A}=10$\ and $\protect\alpha =0.5$, and (b) a site state
at the left end. We plot the imaginary (blue) and real (red) parts of the
amplitudes $\left\langle j\right. \left\vert \protect\phi \left( t\right)
\right\rangle $ of the wave function at instants $t$ with the unit of $1/J$.
We see that in both cases, the evolved wave functions tend to the persistent
emission of the plane wave with the momentum $k\approx k_{c}=\protect\pi /4$.
One can see the plane wave travels to the left and the corresponding phase
velocity is $4\protect\sqrt{2}/\protect\pi $ from the subfigures $t=80$\ and
$t=80.5$ ($t=60$ and $t=60.5$), in the Figs. \protect\ref{fig7} (a) (\protect
\ref{fig7} (b)).}
\label{fig7}
\end{figure*}


\subsection{Current source and drain}

The above results are helpful to understand the mechanism of the Hermiticity
of a non-Hermitian system. It is well known that the existence of the full
real spectra of two above $\mathcal{PT}$\ systems is attributed to\ the
balance between the source and drain. Nevertheless, this description cannot
provide an explanation for the exceptional point and symmetry breaking since
the $\mathcal{PT}$ symmetry of the Hamiltonian seems to maintain such a
balance always. Based on the investigations of above two subsections, we can
reach the following picture: A finite non-Hermitian cluster can act as a
source (or drain), while a semi-infinite lead can act as a tunnel to release
the current caused by the source (drain). A semi-infinite lead has its own
threshold to carry the current, which is characterized by the onset of
complex energy level, or unsteady current. Then the\ exceptional point in
this sense is the threshold of the balance between source (or drain) and the
lead. It leads to another signature of the point,
delocalization-localization transition of the corresponding wave function.

As for a finite $\mathcal{PT}$-symmetric system in the form $H_{\text{1}}$\
and $H_{\text{2}}$, the source and drain is always in balance within the
unbroken region. The consistency of the phase diagrams between the finite $%
\mathcal{PT}$-symmetric system with large $N$\ and the corresponding
semi-infinite systems shows that the exceptional points are caused by the
same mechanism. Then the essence of the symmetry breaking is the unbalance
between the source (drain) and the uniform chain, rather than that between
source and drain. Loosely speaking, the symmetry breaking is due to that the
uniform chain blocks the current from the source and drain. On the other
hand, the accordance between the results of these $\mathcal{PT}$\ systems
with large $N$\ and that of semi-infinite systems implies that a lead can
act as a multifunctional\ source (or drain) to match a drain (or source).

We would like to emphasize the difference between two types of non-Hermitian
elements, complex on-site potential and non-Hermitian hopping amplitude, by
means of the current $\mathcal{J}\left( k\right) =\left\langle k\right\vert
\mathcal{\hat{J}}_{j}\left\vert k\right\rangle $\ for the eigenstate $%
\left\vert k\right\rangle $\ of the Hamiltonian $H_{\text{1}}$\ and $H_{%
\text{2}}$. In Fig. \ref{fig6}, we plot the current $\mathcal{J}\left(
k\right) $\ for all the eigenstates $\left\vert k\right\rangle $\ of
finite-size systems for $H_{\text{1}}$\ and $H_{\text{2}}$. It shows that
the signs of the currents are independent of $k$ for $H_{\text{1}}$, but
dependent of $k$ for $H_{\text{2}}$.\ A certain non-Hermitian dimer can be a
source or drain for two different eigenstates, which is different from a
complex potential.

\section{Wavepacket dynamics}

\label{wavepacket}In this section we will apply the above theoretical
results to simple accessible examples to investigate the dynamic behavior
for local initial states. This may provide some insights into the
application in practice.

Firstly, we focus on the phenomenon of the persistent emission.\ Consider an
arbitrary local initial state on the lead in the system at the exceptional
point. The initial state can always be written in the form
\begin{equation}
\left\langle j\right. \left\vert \phi \left( t=0\right) \right\rangle
=c_{k_{c}}e^{ik_{c}j}+\sum\limits_{k\neq k_{c}}c_{k}\left(
e^{ikj}+R_{k}e^{-ikj}\right) ,
\end{equation}%
where $\sum\nolimits_{k\neq k_{c}}c_{k}\left( e^{ikj}+R_{k}e^{-ikj}\right) $%
\ represents the superposition of the scattering states with different $k$.\
It is presumable that the probability of all the scattering states transfers
to infinity\ after a sufficient long time, and then only the unidirectional
plane wave survives. To demonstrate and verify this analysis, numerical
simulations are performed for two typical initial states: an incoming
Gaussion wavepacket and a delta-pulse at the scattering center. A Gaussian
wavepacket with momentum $k_{0}$ and initial center $N_{A}$ has the form
\begin{equation}
\left\vert \psi (k_{0},N_{A})\right\rangle =\frac{1}{\sqrt{\Omega _{0}}}%
\sum_{j}e^{-\frac{^{\alpha ^{2}}}{2}(j-N_{A})^{2}}e^{ik_{0}j}\left\vert
j\right\rangle ,  \label{GWP}
\end{equation}%
where $\Omega _{0}=\sum_{j}e^{-\alpha ^{2}(j-N_{A})^{2}}$ is the
normalization factor and the half-width of the wavepacket is $2\sqrt{\ln 2}%
/\alpha $. Here we take $k_{0}=-\pi /2$, $N_{A}=100$\ and $\alpha =0.5$. The
concerned system is described by the Hamiltonian in the Eq. (\ref{H_1}) with
$\theta =3\pi /4$, which corresponds to the persistent emission of the plane
wave with momentum $k_{c}=\pi /4$.\ The profiles of the evolved wave
functions are plotted in Figs. \ref{fig7} (a) and \ref{fig7} (b). One can
see the profile of the wave and the corresponding phase velocity from the
figures,\ and after a little long time the evolved wave functions accord
with the plane wave of

\begin{eqnarray}
\text{Re}\left( \left\langle j\right. \left\vert \phi \left( t\right)
\right\rangle \right)  &\sim &\cos \left[ \left( \pi /4\right) j+2Jt\cos
\left( \pi /4\right) \right] , \\
\text{Im}\left( \left\langle j\right. \left\vert \phi \left( t\right)
\right\rangle \right)  &\sim &\sin \left[ \left( \pi /4\right) j+2Jt\cos
\left( \pi /4\right) \right] ,
\end{eqnarray}%
within the finite region along the lead, approximately. This result has
implications in two aspects:\textbf{\ }Firstly, we achieve a better
understanding of the imaginary potential. We found that a complex potential
always corresponds to the wave vector of the unidirectional plane wave,
which is determined by Eq. (\ref{potentialk}).\textbf{\ }Secondly, it
provides a way\textbf{\ }to measure the complex potential in the experiment.
\begin{figure*}[tbp]
\includegraphics[ bb=20 182 535 613, width=6.5 cm, clip]{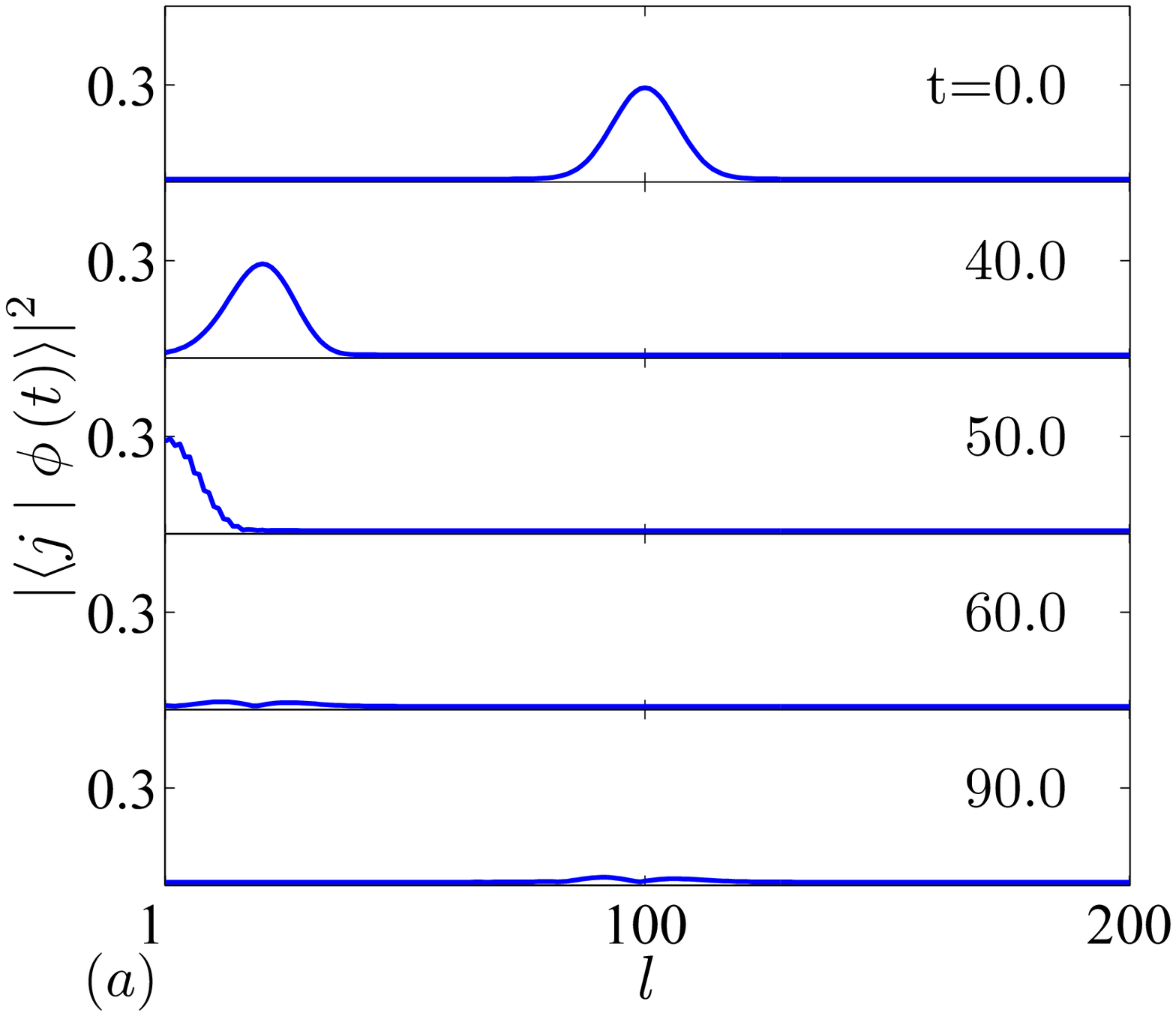} %
\includegraphics[ bb=2 182 517 613, width=6.5 cm, clip]{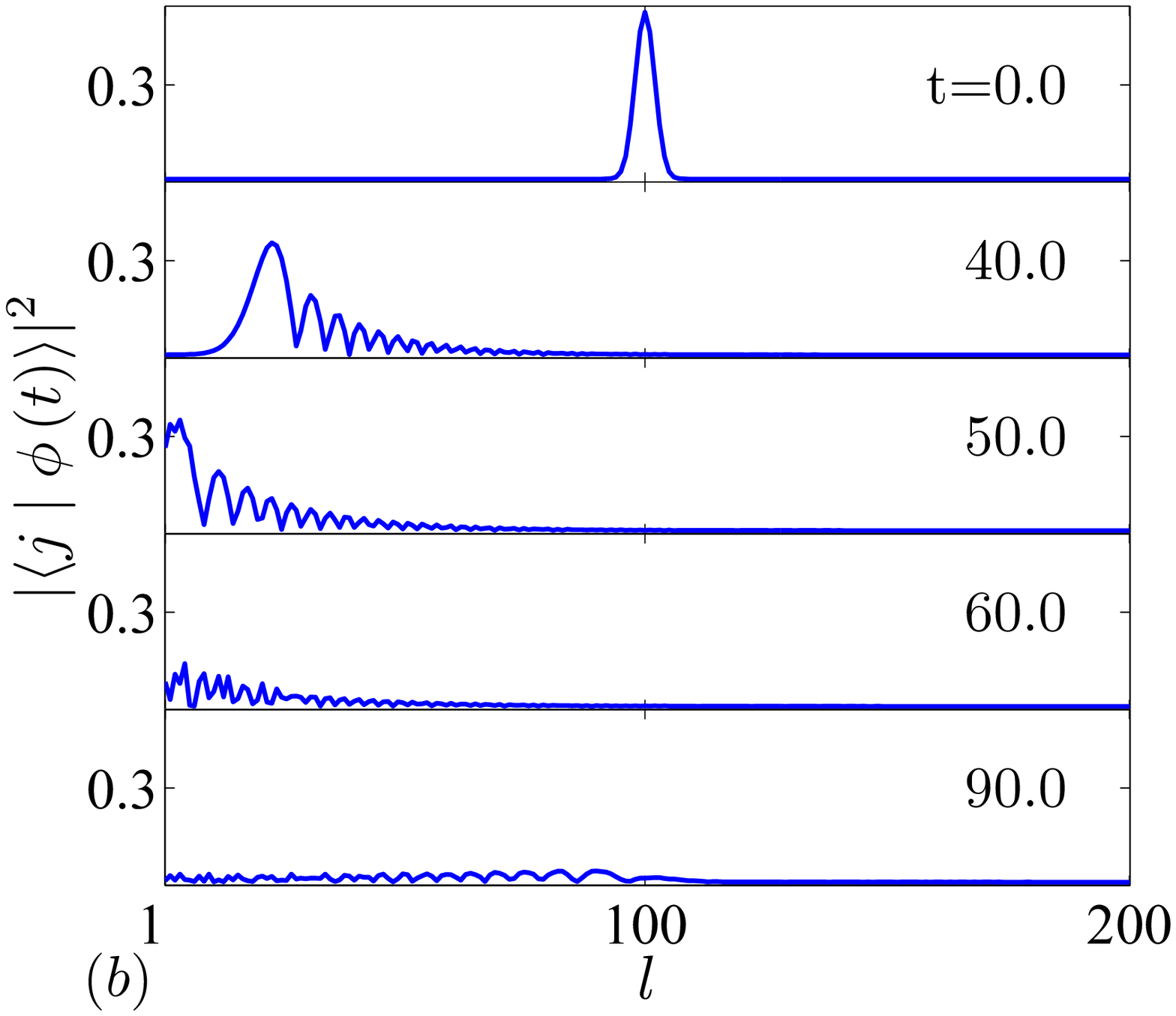}
\caption{(Color online) The probability distribution of the evolved wave
function for the initial state being the incident Gaussian wavepackets with $%
k_{0}=-\protect\pi /2$, $N_{A}=100$, (a) $\protect\alpha =0.5$, and (b) $%
\protect\alpha =0.015$. We plot $\left\vert \left\langle j\right. \left\vert
\protect\phi \left( t\right) \right\rangle \right\vert ^{2}$ at instants $t$
with the unit of $1/J$. We see that in both cases, the reflection
coefficients are small, especially for the wider incident wavepacket, which
is more close to the plane wave with the momentum $k=k_{c}=\protect\pi /2$.}
\label{fig8}
\end{figure*}


On the other hand, it is presumable that the reflectionless absorbtion
should naturally be reflected in the dynamics of the wavepacket with the
momenta around $k_{c}$ due to the continuity of the reflection coefficient
in the vicinity of the exceptional point $k_{c}$. Consider an incoming
Gaussian wavepacket with momentum $k_{0}=k_{c}$ and initial center $N_{A}$,
which can always be written as
\begin{equation}
\left\langle j\right. \left\vert \psi (k_{c},N_{A})\right\rangle =\frac{1}{%
\sqrt{\Omega }}\int_{-\pi }^{\pi }e^{-\frac{1}{2\alpha ^{2}}\left(
k-k_{c}\right) ^{2}-iN_{c}\left( k-k_{c}\right) }\text{d}k\left\langle
j\right. \left\vert k\right\rangle ,  \label{GWP(k)}
\end{equation}%
where $\left\langle j\right. \left\vert k\right\rangle =e^{ikj}$ denotes the
plane wave with momentum $k$, and \ $\Omega =\int_{-\pi }^{\pi }e^{-\left(
k-k_{c}\right) ^{2}/2\alpha ^{2}}$d$k$ is the normalization factor. It is a
superposition of the plane waves with momenta around $k_{c}$, which have
small reflection coefficients. Therefore there is small probability being
reflected for the wavepacket. The optimal situations to achieve the
reflectionless absorbtion of a wavpacket are determined by the conditions%
\begin{equation}
\left\vert R_{k_{c}}\right\vert =0\text{, }\left. \frac{\text{d}\left\vert
R_{k}\right\vert }{\text{d}k}\right\vert _{k=k_{c}}=0,
\end{equation}%
which minimizes the reflectional probability.\ Straightforward derivation
indicates that the optimal complex-potential (hopping) scattering center
system requires $\theta =\pi /2$\ ($\kappa ^{2}=1\pm i$) and the momentum of
the corresponding reflectionless wave is $k_{c}=\pi /2$ ($\pi /4$).

To demonstrate and verify this analysis, numerical simulations are performed
for two initial wavepackets with $k_{0}=-\pi /2$, $N_{A}=100$, $\alpha =0.5$
and $0.15$, respectively. The concerned system is described by the
Hamiltonian in the Eq. (\ref{H_1}) with $\theta =\pi /2$, which corresponds
to the reflectionless plane wave with momentum $k_{c}=\pi /2$.\ The profiles
of the evolved wave functions are plotted in Figs. \ref{fig8} (a) and \ref%
{fig8} (b). The reflection coefficients of the two wavepackets are $0.003$
and $0.035$, respectively. It shows that wider wavepacket leading to lower
reflection rate, which accords with the previous theoretical analysis.

\section{Summary}

\label{sec_summary}

In summary, the mechanism of the non-Hermiticity of a discrete non-Hermitian
system has been investigated in an alternative way. It is shown that the
symmetry is not the necessary condition for the occurrence of full real
spectrum. The underlying mechanism can be explained as the balance between a
non-Hermitian cluster and a semi-infinite chain: the semi-infinite lead can
play a complete role to balance a finite non-Hermitian cluster, resulting in
a full real spectrum. It is also shown that the threshold of such balance is
the exceptional point of the semi-infinite non-Hermitian systems, the
occurrence of the corresponding complex eigenstates experienced the
delocalization-localization transition. Furthermore, at the exceptional
point, the eigen wave function is shown to be a unidirectional plane wave.
Practical application of this feature to the dynamics of the wave packet
demonstrates the phenomena of the self-sustained emission and reflectionless
absorbtion,\ which could be very useful for the design of quantum devices.

\acknowledgments We acknowledge the support of National Basic Research
Program (973 Program) of China under Grant No. 2012CB921900.


\begin{thebibliography}{99}
\bibitem{Scholtz} F. G. Scholtz, H. B. Geyer, and F. J.W.Hahne, Ann. Phys.
(NY) \textbf{213}, 74 (1992).

\bibitem{Bender 98} C. M. Bender, and S. Boettcher, Phys. Rev. Lett. \textbf{%
80}, 5243 (1998).

\bibitem{Bender 99} C. M. Bender, S. Boettcher, and P. N. Meisinger, J.
Math. Phys. \textbf{40}, 2201 (1999).

\bibitem{Dorey 01} P. Dorey, C. Dunning, and R. Tateo, J. Phys. A: Math.
Gen. \textbf{34}, L391 (2001); P. Dorey, C. Dunning, and R. Tateo, J. Phys.
A: Math. Gen. \textbf{34}, 5679 (2001).

\bibitem{Bender 02} C. M. Bender, D. C. Brody, and H. F. Jones, Phys. Rev.
Lett. \textbf{89}, 270401 (2002).

\bibitem{A.M43} A. Mostafazadeh, J. Math. Phys. \textbf{43,} 205 (2002); J.
Math. Phys. \textbf{43,} 2814 (2002); J. Math. Phys. \textbf{43}, 3944
(2002);

\bibitem{A.M36} A. Mostafazadeh, J. Phys. A: Math. Gen. \textbf{36}, 7081
(2003).

\bibitem{A.M} A. Mostafazadeh and A. Batal, J. Phys. A: Math. Gen. \textbf{37%
}, 11645 (2004).

\bibitem{Jones} H. F. Jones, J. Phys. A: Math. Gen. \textbf{38}, 1741 (2005).

\bibitem{ZXZ} X. Z. Zhang and Z. Song, Phys. Rev. A \textbf{87}, 012114
(2013).

\bibitem{Ruschhaupt} A Ruschhaupt, F Delgado and J G Muga, J. Phys. A: Math.
Gen. \textbf{38,} L171 (2005).

\bibitem{R. El-Ganainy} R. El-Ganainy, K. G. Makris, D. N. Christodoulides,
and Z. H. Musslimani, Opt. Lett. \textbf{32}, 2632 (2007).

\bibitem{K. G. Makris} K. G. Makris, R. El-Ganainy, D. N. Christodoulides,
and Z. H. Musslimani, Phys. Rev. Lett. \textbf{100}, 103904 (2008).

\bibitem{Christodoulides} K. G. Makris, R. El-Ganainy, D. N.
Christodoulides, and Z. H. Musslimani, Phys. Rev. A \textbf{81}, 063807
(2010).

\bibitem{Z. H. Musslimani} Z. H. Musslimani, Phys. Rev. Lett. \textbf{100},
030402 (2008).

\bibitem{S. Klaiman} S. Klaiman, U. G\"{u}nther, and N. Moiseyev, Phys. Rev.
Lett. \textbf{101}, 080402 (2008).

\bibitem{S. Longhi} S. Longhi, Phys. Rev. Lett. \textbf{103}, 123601 (2009).

\bibitem{H. Schomerus} H. Schomerus, Phys. Rev. Lett. \textbf{104}, 233601
(2010).

\bibitem{LonghiLaser} S. Longhi, Phys. Rev. A \textbf{82}, 031801(R) (2010);
Phys. Rev. Lett. \textbf{105}, 013903 (2010).

\bibitem{YDChong} Y. D. Chong, Li Ge, Hui Cao and A. D. Stone, Phys. Rev.
Lett. \textbf{105}, 053901 (2010).

\bibitem{Keya} K. Zhou, Z. Guo, J. Wang and S. Liu Opt. Lett. \textbf{35},
2928 (2010).

\bibitem{Guo} A. Guo \textit{et al.}, Phys. Rev. Lett. \textbf{103}, 093902
(2009).

\bibitem{Kottos} C. E. R\"{u}ter \textit{et al.}, Nat. Phys. \textbf{6}, 192
(2010); T. Kottos, ibid. \textbf{6}, 166 (2010); A. Regensburger \textit{et
al.}, Nature \textbf{488}, 167 (2012).

\bibitem{Znojil} M. Znojil, Phys. Lett. B \textbf{650}, 440 (2007); J. Phys.
A \textbf{40}, 13131 (2007); Phys. Rev. A \textbf{82}, 052113 (2010).

\bibitem{a1} M. Znojil, J. Math. Phys. \textbf{50}, 122105 (2009); J. Phys.
A: Math. Theor. \textbf{44}, 075302 (2011), Phys. Lett. A \textbf{375}, 2503
(2011); M. Znojil and M. Tater, Int. J. Theor. Phys. \textbf{50}, 982 (2011).

\bibitem{Bendix} O. Bendix, R. Fleischmann, T. Kottos and B. Shapiro, Phys.
Rev. Lett. \textbf{103}, 030402 (2009).

\bibitem{Liang Jin} L. Jin and Z. Song, Phys. Rev. A \textbf{80}, 052107
(2009); Phys. Rev. A \textbf{81}, 032109 (2010).

\bibitem{Longhi} S. Longhi, Phys. Rev. B \textbf{82}, 041106(R) (2010).

\bibitem{Joglekar} Y. N. Joglekar, D. Scott, M. Babbey, and A. Saxena, Phys.
Rev. A \textbf{82}, 030103(R) (2010).

\bibitem{Joglekar1} Y. N. Joglekar and A. Saxena, Phys. Rev. A \textbf{83},
050101(R) (2011); D. D. Scott and Y. N. Joglekar, Phys. Rev. A \textbf{83},
050102(R) (2011); Y. N. Joglekar and J. L. Barnett, Phy. Rev. A \textbf{84},
024103 (2011).

\bibitem{Joglekar2} D. D. Scott and Y. N. Joglekar Phys. Rev. A \textbf{85},
062105 (2012); J. Phys. A: Math. Theor. \textbf{45}, 444030 (2012).

\bibitem{Korff} C. Korff and R. Weston, J. Phys. A \textbf{40}, 8845 (2007);
O. A. Castro-Alvaredo and A. Fring, \textit{ibid}. \textbf{42}, 465211
(2009).

\bibitem{T. Deguchi} T. Deguchi and P. K. Ghosh, J. Phys. A: Math. Theor.
\textbf{42}, 475208 (2009).

\bibitem{Giorgi} G. L. Giorgi, Phys. Rev. B \textbf{82}, 052404 (2010).

\bibitem{ZXZ1} X. Z. Zhang, L. Jin, and Z. Song, Phys. Rev. A \textbf{85},
012106 (2012).

\bibitem{H. Zhong} H. Zhong, W. Hai, G. Lu, and Z. Li, Phys. Rev. \textbf{A}
84, 013410 (2011).

\bibitem{L. Jin} L. Jin and Z. Song, Ann. Phys. \textbf{330}, 142 (2013).

\bibitem{Stefano} S. Longhi, Phy. Rev. B \textbf{81}, 195118 (2010); Phy.
Rev. A \textbf{82}, 032111 (2010).

\bibitem{Znojil2008} M. Znojil, J. Phys. A: Math. Theor. \textbf{41}, 292002
(2008).

\bibitem{a2} L. Jin and Z. Song, Phys. Rev. A \textbf{81}, 022107 (2010);
Phys. Rev. A \textbf{84}, 042116 (2011).

\bibitem{a3} L. Jin and Z. Song, Phys. Rev. A \textbf{85}, 012111 (2012);\
W. H. Hu, L. Jin, Y. Li, and Z. Song, Phys. Rev. A \textbf{86}, 042110
(2012).

\bibitem{a4} A. Regensburger, C. Bersch, M.-A. Miri, G. Onishchukov, D. N.
Christodoulides and U. Peschel, Nature \textbf{488}, 167 (2012); M.-A. Miri,
A. Regensburger, U. Peschel, D. N. Christodoulides, Phys. Rev. A \textbf{86}%
, 023807 (2012).

\bibitem{Caroli} C. Caroli, R. Combescot, P. Nozieres, and D. Saint-James,
J. Phys. C: Solid St. Phys. \textbf{20}, 1018 (1965).
\end{thebibliography}
\end{document}